\documentclass[aps,prd,twocolumn]{revtex4-1}
\usepackage[colorlinks=true, linkcolor=blue, citecolor=blue, urlcolor=blue]{hyperref}
\usepackage{bm,latexsym,amssymb,amsmath,mathrsfs}
\usepackage{graphicx}
\usepackage{color}
\usepackage{times,setspace}

\newcommand{\re}{{\rm Re}}

\newcommand{\hmu}{\hat{\mu}}

\begin{document}

\title{'t Hooft surface in lattice gauge theory}
\author{Takuya Shimazaki and Arata Yamamoto}
\affiliation{Department of Physics, The University of Tokyo, Tokyo 113-0033, Japan}

\begin{abstract}
We discuss the lattice formulation of the 't Hooft surface, that is, the two-dimensional surface operator of a dual variable.
The 't Hooft surface describes the world sheets of topological vortices.
We derive the formulas to calculate the expectation value of the 't Hooft surface in the multiple-charge lattice Abelian Higgs model and in the lattice non-Abelian Higgs model.
As the first demonstration of the formula, we compute the intervortex potential in the charge-2 lattice Abelian Higgs model.
\end{abstract}

\maketitle

\section{Introduction}

Topological orders are quantum phases beyond the Landau theory of symmetry breaking \cite{Wen:2004ym}.
Local order parameters fail to capture their manifestation.
Some topological orders are characterized by long-range quantum entanglements exhibiting fractional excitations and topological degeneracy~\cite{Laughlin:1983fy,Arovas:1984qr,Wen:1989zg,Wen:1990zza,Kitaev:1997wr,Chen:2010gda,Savary:2016ksw}.
One of the well-established approaches is a higher-form symmetry~\cite{Gaiotto:2014kfa,Hofman:2018lfz,Wen:2018zux,Komargodski:2017dmc}.
Order parameters for the generalized global symmetries are extended objects, e.g., loop, surface, and so on.
Their topological natures make it possible to classify the topological orders.

The most famous example of the non-local order parameters is a loop operator.
For instance, in SU($N$) gauge theory, the Wilson loop \cite{Wilson:1974sk} is defined by the path-ordered product of the loop integral
\begin{equation}
  W[\mathcal{C}] = P e^{i\oint_\mathcal{C} A_\mu(x) dx^\mu}.
\end{equation}
Since the gauge field $A_\mu$ couples to electrically charged particles, the Wilson loop describes the world lines of charged particles.
The SU($N$) gauge theory also has magnetically charged particles, namely, magnetic monopoles.
The world lines of magnetic monopoles define the 't Hooft loop~\cite{tHooft:1977nqb}.
Using the dual field $\tilde{A}_\mu$, the 't Hooft loop is given by
\begin{equation}
  \tilde{W}[\mathcal{C}^*] = e^{i\oint_{\mathcal{C}^*} \tilde{A}_\mu(x) dx^\mu}.
\end{equation}
The Wilson and 't Hooft loops are the order parameters for the confinement of electric and magnetic charges, respectively.

A more nontrivial example is a surface operator.
In the gauge theory coupled to Higgs fields, topological vortices appear.
The vortices are one-dimensional defects, so their trajectories form world sheets.
The vortex world sheets are two-dimensional defects with delta function support on the surfaces.
The surface operator to create the vortex world sheet is referred to as the {\it 't Hooft surface} \cite{Note1}.
For instance, in the $BF$ theory, it is given by the closed-surface integral over the dual 2-form field $B_{\mu\nu}$,
\begin{equation}
  \tilde{V}[\mathcal{S}^*] = e^{i\oint_{\mathcal{S}^*} B_{\mu\nu} dS^{\mu\nu}}\, .
\end{equation}
When the theory has Z$_N$ topological order, the vortices have fractional magnetic charge $q/Ne$ ($q\in \mathbb{Z}$).
The 't Hooft surface gives the criterion for the confinement of the fractionally charged vortices.
It plays an essential role in the topological order of Cooper pairs in superconductors \cite{Hansson_2004,Diamantini:2011gi,PhysRevB.90.085114} and  that of diquarks in color superconductors \cite{Nishida:2010wr,Cherman:2018jir,Hirono:2018fjr,Hidaka:2019jtv,Hirono:2019oup}.

These 't Hooft operators have crucial difficulty.
The 't Hooft operators can be elegantly formulated in topological quantum field theory, that is, effective theory to reproduce topological properties of the original quantum field theory.
The calculation beyond the effective theory is, however, not easy.
The dual gauge fields are not fundamental fields but defects or singularities in the original theory.
It might seem impossible to calculate the expectation values of the 't Hooft operators in a quantitative manner.
Surprisingly, this is possible in lattice gauge theory.
The formulation has been known for the 't Hooft loop in the Yang-Mills theory \cite{Mack:1978kr,Ukawa:1979yv,Srednicki:1980gb}.
It was applied to several lattice simulations \cite{Billoire:1981ye,DeGrand:1981yx,Kovacs:2000sy,Hart:2000hq,Hoelbling:2000su,DelDebbio:2000cb,DelDebbio:2000cx,deForcrand:2000fi,deForcrand:2001nd,deForcrand:2005pb,deForcrand:2005zg}.
This can be generalized to higher-dimensional defects, say, the 't Hooft surface.

In this paper, we study the 't Hooft surface in lattice gauge theory.
After introducing the basics of dual variables in Sec.~\ref{secdual}, we discuss how to compute the 't Hooft surface in lattice simulation.
We focus on the lattice gauge Higgs models with Z$_N$ topological order: the Abelian Higgs model in Sec.~\ref{secAHmodel} and the non-Abelian Higgs model in Sec.~\ref{secNAHmodel}.
The simulation results are shown in Sec.~\ref{secsimulation}.
Finally, Sec.~\ref{secsum} is devoted to summary.
The Euclidean four-dimensional lattice is considered and the lattice unit is used throughout the paper.

\section{Dual lattice and dual variable}
\label{secdual}

\begin{figure}[h]
\begin{center}
 \includegraphics[width=.25\textwidth]{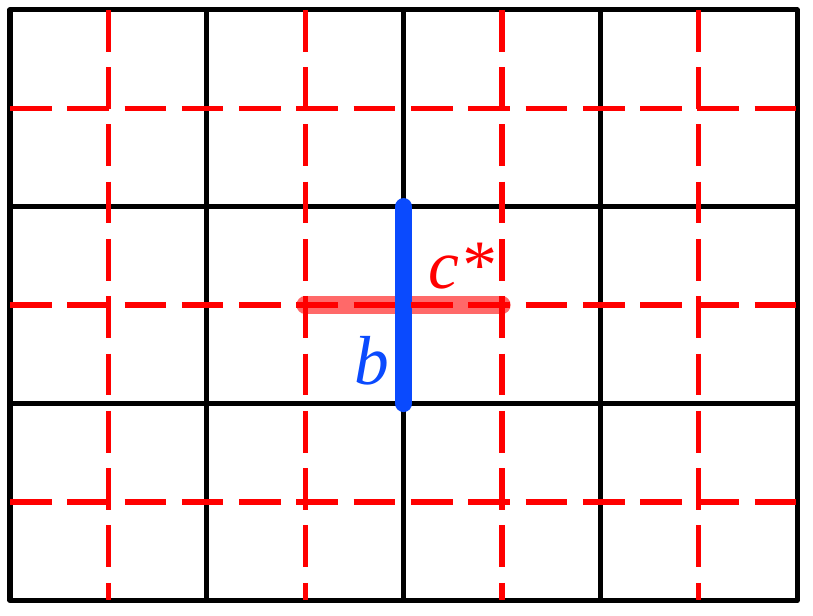}
\caption{
\label{figbtoc}
Original lattice (black solid lines) and dual lattice (red broken lines).
In four dimensions, the bond $b$ (blue thick line) is dual to the dual cube $c^*$ (red thick line).
}
\end{center}
\end{figure}

Dual variables live on the dual lattice.
The dual lattice is defined by translating the original lattice by a half of lattice spacing in all directions.
The schematic figure is shown in Fig.~\ref{figbtoc}.
The sites of the dual lattice are in positions of the hypercube centers of the original lattice.
In four dimensions, there are the one-to-one correspondences between $d$-dimensional objects on the original lattice and $(4-d)$-dimensional objects on the dual lattice: a bond $b$ is dual to a cube $c^*$, a plaquette $p$ is dual to a plaquette $p^*$, etc.
The asterisks denote the objects on the dual lattice.

In lattice gauge theory, there are two ways to introduce dual variables.
One is to replace all the integral variables in the path integral by their dual variables.
The path integral is completely reformed.
For example, the gauge Higgs theory is described by dual plaquette variables and dual cube variables \cite{Hayata:2019rem} (see also Refs.~\cite{Gattringer:2012jt,Mercado:2013yta,Mercado:2013ola,Gattringer:2018dlw,Goschl:2018uma}).
The other is to insert the dual variables without reforming the path integral.
For example, when a $\pi$-flux exists on $p^*$, the plaquette variable on $p$ changes as $U_{\mu\nu} \to e^{i\pi} U_{\mu\nu}$ but the integral variables do not change.
This is exactly what is done in the lattice formulation of the 't Hooft loop \cite{Mack:1978kr,Ukawa:1979yv,Srednicki:1980gb}.
We consider such insertion of the 't Hooft surface in the following sections.

\section{Abelian Higgs model}
\label{secAHmodel}

We first consider Abelian gauge theory.
When the Higgs field has the multiple electric charge $Ne$ $(N\in \mathbb{Z})$, the theory has Z$_N$ topological order.
The magnetic charge of a vortex is fractionally quantized to be $1/Ne$.
The theory is often referred as the charge-$N$ lattice Abelian Higgs model \cite{Fradkin:1978dv}.
The path integral is given by
\begin{equation}
 Z_0 = \int DA D\phi \ e^{-S}
\end{equation}
with the Abelian gauge field $A_\mu$ and the complex scalar field $\phi$.
Let us define the link variable
\begin{equation}
 U_{\mu}(x) = e^{iA_\mu(x)}
\end{equation}
and the plaquette variable
\begin{equation}
 U_{\mu\nu}(x) = U_\mu(x)U_\nu(x+\hmu)U^{-1}_\mu(x+\hat\nu)U^{-1}_\nu(x)
,
\end{equation}
where $\hat{\mu}$ stands for the unit lattice vector in the $\mu$ direction.
The Euclidean action is given by
\begin{equation}
 S = S_{\rm gauge} + S_{\rm local} + S_{\rm hop}
\end{equation}
with the gauge part
\begin{equation}
S_{\rm gauge} = - \frac{1}{2e^2} \sum_{x} \sum_{\mu,\nu} \re U_{\mu\nu}(x)
,
\end{equation}
the local part of the scalar field
\begin{equation}
S_{\rm local} = \sum_x \left[ 8 \phi^*(x)\phi(x) + \lambda \left\{ \phi^*(x)\phi(x) - v^2 \right\}^2  \right]
,
\end{equation}
and the hopping part
\begin{equation}
\begin{split}
\label{eqShop}
S_{\rm hop} =& - \sum_{x} \sum_{\mu} \bigg\{ \phi^*(x) U_\mu^N(x) \phi(x+\hat{\mu})
\\
& + \phi^*(x+\hat{\mu}) U_\mu^{-N}(x) \phi(x) \bigg\}
.
\end{split}
\end{equation}
The theory is invariant under the local U(1) gauge transformation
\begin{eqnarray}
 U_\mu(x) &\to&  \Lambda(x) U_\mu(x) \Lambda^*(x+\hmu)
\\
 \phi(x) &\to& \Lambda^N(x) \phi(x)
\end{eqnarray}
and the global $Z_N$ transformation
\begin{equation}
 U_\mu(x) \to U_\mu(x) e^{i2\pi/N}.
\end{equation}
Because of the Z$_N$ symmetry, the vacuum is $N$-fold degenerate.

\begin{figure}[h]
\begin{center}
 \includegraphics[width=.3\textwidth]{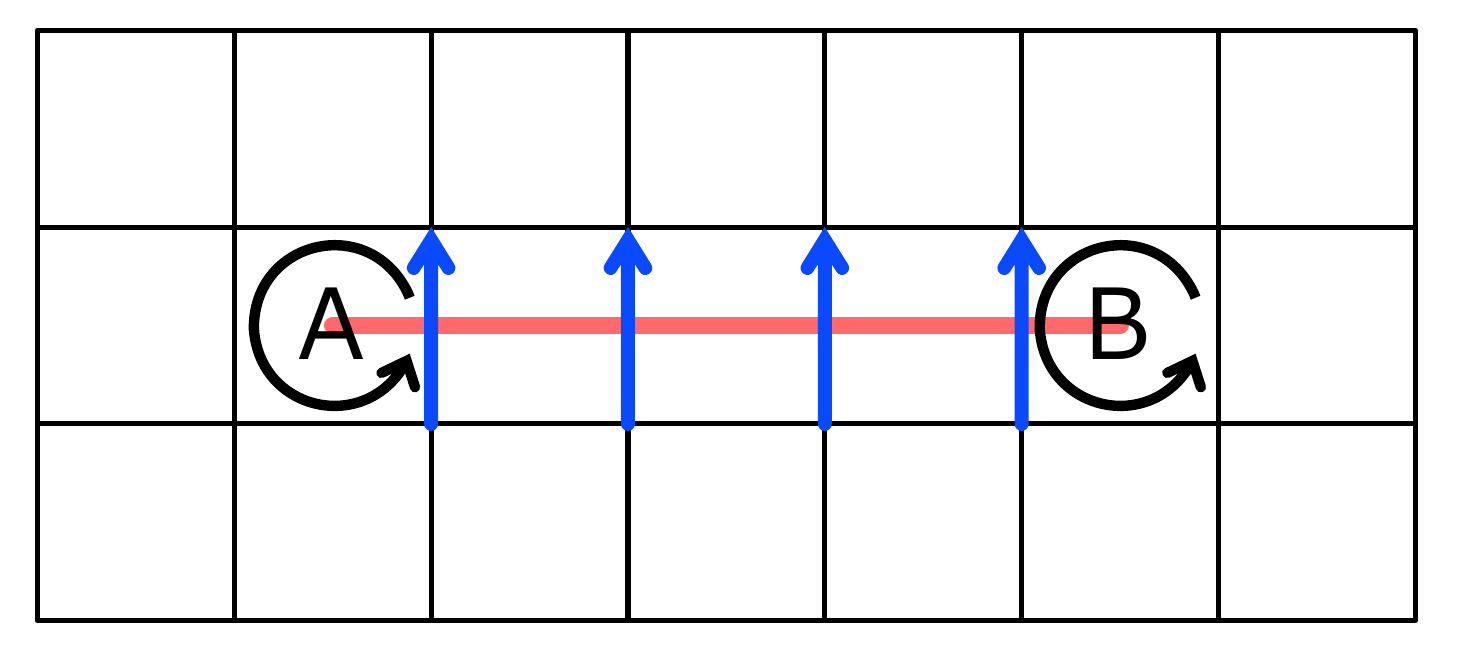}
\caption{
\label{figplane}
Schematic figure for Eq.~\eqref{eqSprime}.
The 't Hooft surface is inserted on the three-dimensional volume $\mathcal{V}^*$ (red thick line).
The Z$_N$ element $e^{i2\pi q/N}$ is multiplied to the hopping terms penetrating $\mathcal{V}^*$ (blue arrows).
}
\end{center}
\end{figure}

We put the 't Hooft surface of magnetic charge $q/Ne$ on a two-dimensional closed surface $\mathcal{S}^*$.
The path integral changes as
\begin{eqnarray}
\label{eqZprime}
 Z_{\mathcal{S}^*} &=& \int DA D\phi \ e^{-S'}
\\
 S' &=& S_{\rm gauge} + S_{\rm local} + S'_{\rm hop}
.
\end{eqnarray}
The hopping part is modified as
\begin{equation}
\begin{split}
\label{eqSprime}
S'_{\rm hop} =& - \sum_{x,\mu\in B(\mathcal{V}^*)} \bigg\{ e^{i2\pi q/N} \phi^*(x) U_\mu^N(x) \phi(x+\hat{\mu})
\\
& + e^{-i2\pi q/N} \phi^*(x+\hat{\mu}) U_\mu^{-N}(x) \phi(x) \bigg\}
\\
& - \sum_{x,\mu\notin  B(\mathcal{V}^*)} \bigg\{ \phi^*(x) U_\mu^N(x) \phi(x+\hat{\mu})
\\
& + \phi^*(x+\hat{\mu}) U_\mu^{-N}(x) \phi(x) \bigg\}
,
\end{split}
\end{equation}
where $B(\mathcal{V}^*)$ is defined by the hopping terms penetrating $\mathcal{V}^*$, s.t.~ $\mathcal{S}^* = \partial \mathcal{V}^*$ (see Fig.~\ref{figplane}).
Therefore, the expectation value of the 't Hooft surface is given by the formula
\begin{equation}
\label{eqexW}
 \langle \tilde{V}[\mathcal{S}^*] \rangle = \frac{Z_{\mathcal{S}^*}}{Z_0} = \frac{\int DA D\phi \ e^{-\Delta S} e^{-S}}{\int DA D\phi \ e^{-S}} = \langle e^{-\Delta S} \rangle
\end{equation}
with
\begin{equation}
\begin{split}
\label{eqDeltaS}
\Delta S =& \ S' - S = S'_{\rm hop} - S_{\rm hop}
\\
=& - \sum_{x,\mu\in B(\mathcal{V}^*)} \bigg\{ \left( e^{i2\pi q/N} -1 \right) \phi^*(x) U_\mu^N(x) \phi(x+\hat{\mu})
\\
& + \left( e^{-i2\pi q/N} - 1 \right) \phi^*(x+\hat{\mu}) U_\mu^{-N}(x) \phi(x) \bigg\}
.
\end{split}
\end{equation}
This formula is the main result of this paper.
The explicit derivation is given in Appendix \ref{appder}.

The above formula can be intuitively understood in Fig.~\ref{figplane}.
The red line is the three-dimensional volume $\mathcal{V}^*$ on the dual lattice.
The hopping terms penetrating $\mathcal{V}^*$ are multiplied by the Z$_N$ element $e^{i2\pi q/N}$.
The winding number of each plaquette is given by the sum of the angles of the four hopping terms.
As shown by the circle arrows in the figure, the Z$_N$ element changes the winding number by $+q/N$ at A, and by $-q/N$ at B, and by 0 elsewhere.
This means that a vortex and an antivortex are inserted at $\mathcal{S}^* = \partial \mathcal{V}^*$.

There are two remarks on the above formula.
The first one is that the inserted vortices have fractional winding numbers.
They are different from the standard vortices defined by integer winding numbers.
The above formula cannot realize the integer winding numbers because $\Delta S = 0$ for $q/N\in \mathbb{Z}$.
Another formulation is necessary to insert the vortices with integer winding numbers \cite{Yamamoto:2020qxj}.
The second one is that $\mathcal{V}^*$ is non-unique.
In four dimensions, $\mathcal{V}^*$ can be deformed in the perpendicular direction by integral variable transformation, as long as $\mathcal{S}^*$ is fixed.
The same ambiguity exists in the 't Hooft loop \cite{Kovacs:2000sy}.
On the other hand, $\mathcal{S}^*$ is invariant under the transformation.
This means that the positions of the vortices are physical.

\section{Non-Abelian Higgs model}
\label{secNAHmodel}

Next let us consider non-Abelian gauge theory.
Among the non-Abelian gauge Higgs models, the $N$-color and $N$-flavor case is of special importance.
When the numbers of color and flavor are equal, the non-Abelian vortex, as well as the Abelian vortex, can exist.
The minimal winding number of the non-Abelian vortex is $1/N$, while the winding number of the Abelian vortex is integer.

In the $N$-color and $N$-flavor non-Abelian Higgs model, the hopping part of the lattice action is
\begin{equation}
\begin{split}
S_{\rm hop} =& - \sum_x \sum_\mu \sum_i \bigg\{ \phi_i^\dagger(x) U_\mu(x) \phi_i(x+\hat{\mu})
\\
& + \phi_i^\dagger(x+\hat{\mu}) U_\mu^{-1}(x) \phi_i(x) \bigg\}
.
\end{split}
\end{equation}
This is is almost the same as Eq.~\eqref{eqShop}, except that the link variable $U_\mu$ is a U($N$) element and the $N$-flavor scalar fields $\phi_i \, (i=1,\cdots,N)$ are $N$-component vectors.
The other parts are quite different, but they are irrelevant for the present argument.
(See Ref.~\cite{Yamamoto:2018vgg} for the complete form of the lattice action.)
We can easily derive the formula; Eq.~\eqref{eqDeltaS} is replaced by
\begin{equation}
\begin{split}
\Delta S =& - \sum_{x,\mu\in B(\mathcal{V}^*)} \bigg\{ \left( e^{i2\pi q/N} -1 \right) \phi^\dagger(x) U_\mu(x) \phi(x+\hat{\mu})
\\
& + \left( e^{-i2\pi q/N} - 1 \right) \phi^\dagger(x+\hat{\mu}) U_\mu^{-1}(x) \phi(x) \bigg\}
.
\end{split}
\end{equation}
The inserted 't Hooft surface satisfies the same properties as in the Abelian case.

\section{Simulation}
\label{secsimulation}

To demonstrate the above formalism, we perform the numerical simulation in the simplest case, the charge-$2$ lattice Abelian Higgs model.
The simulation details are summarized in Appendix \ref{appsim}.
The Z$_2$ group has two elements: one trivial state $+1$ (zero magnetic charge) and one nontrivial state $-1$ (fractional magnetic charge $1/2e$).
This means that a vortex and an antivortex are equivalent.
This specialty is due to the compactness of the link variable.
In general, the lattice Abelian Higgs model sometimes shows different behaviors from the continuous Abelian Higgs model.
The results should be interpreted as the properties of lattice superconductors, not of realistic superconductors in continuum space.

\begin{figure}[h]
\begin{center}
 \includegraphics[width=.48\textwidth]{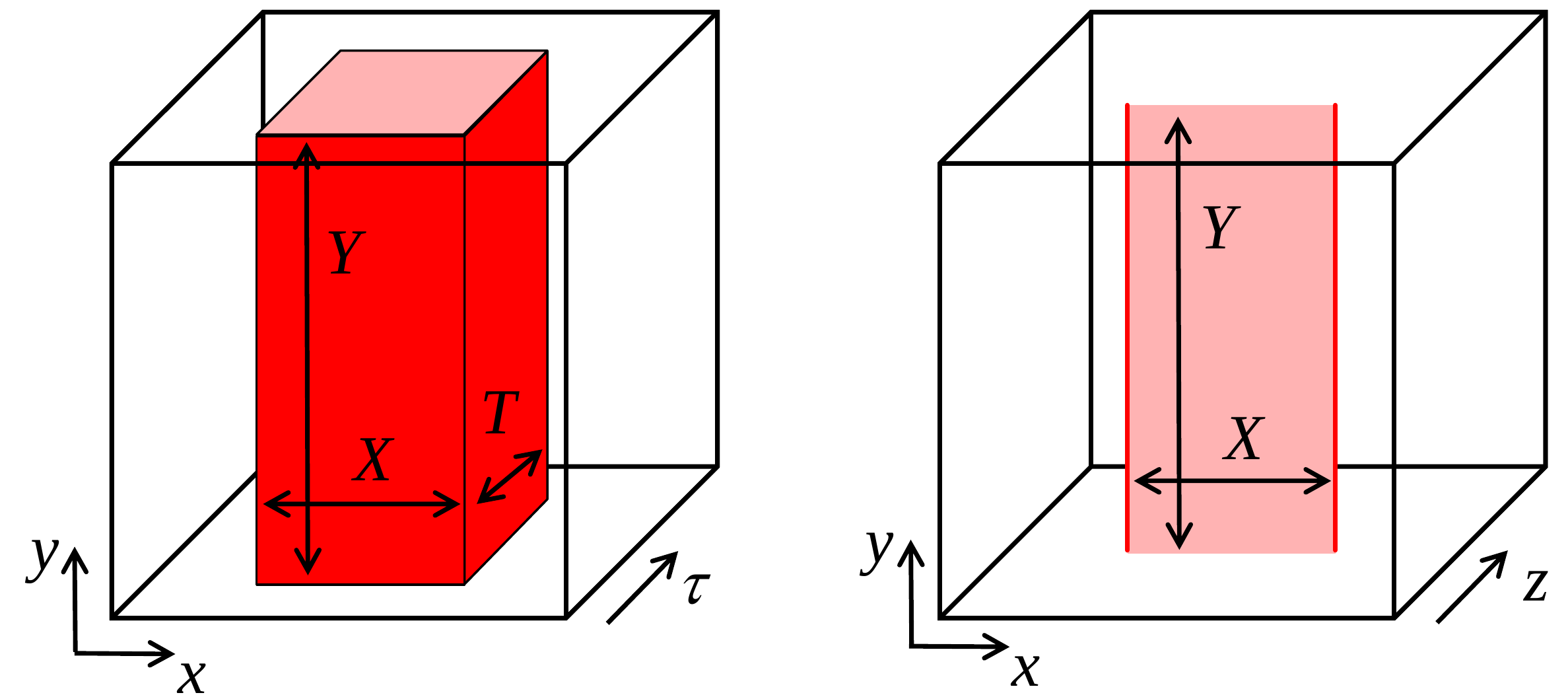}
\caption{
\label{figbox}
Geometry projected on the $xy\tau$ hyperplane (left) and on the $xyz$ hyperplane (right).
The two-dimensional 't Hooft surface (dark red) surrounds the three-dimensional volume $\mathcal{V}^*$ (light red).
}
\end{center}
\end{figure}

We consider the four-dimensional hypercuboid $N_x \times N_y \times N_z \times N_\tau$ with periodic boundary conditions.
As shown in Fig.~\ref{figbox}, the 't Hooft surface is inserted on the dual cuboid $X \times Y \times T$ inside the hypercuboid.
We take $Y=N_y$ to simplify the analysis.
The nontrivial Z$_2$ element $-1$ is multiplied to the hopping terms in the $z$ direction.
When the time extent $T$ is large enough, the 't Hooft surface asymptotically behaves as
\begin{equation}
\label{eqET}
 \langle \tilde{V}[\mathcal{S}^*] \rangle \propto e^{-ET}
,
\end{equation}
where $E$ is the energy of a static and straight vortex-antivortex pair.
Since $E$ is proportional to $Y$ because of translational invariance, $E/Y$ is a function of $X$.
Therefore, $E/Y$ can be interpreted the intervortex potential per length.

\begin{figure}[h]
\begin{center}
 \includegraphics[width=.48\textwidth]{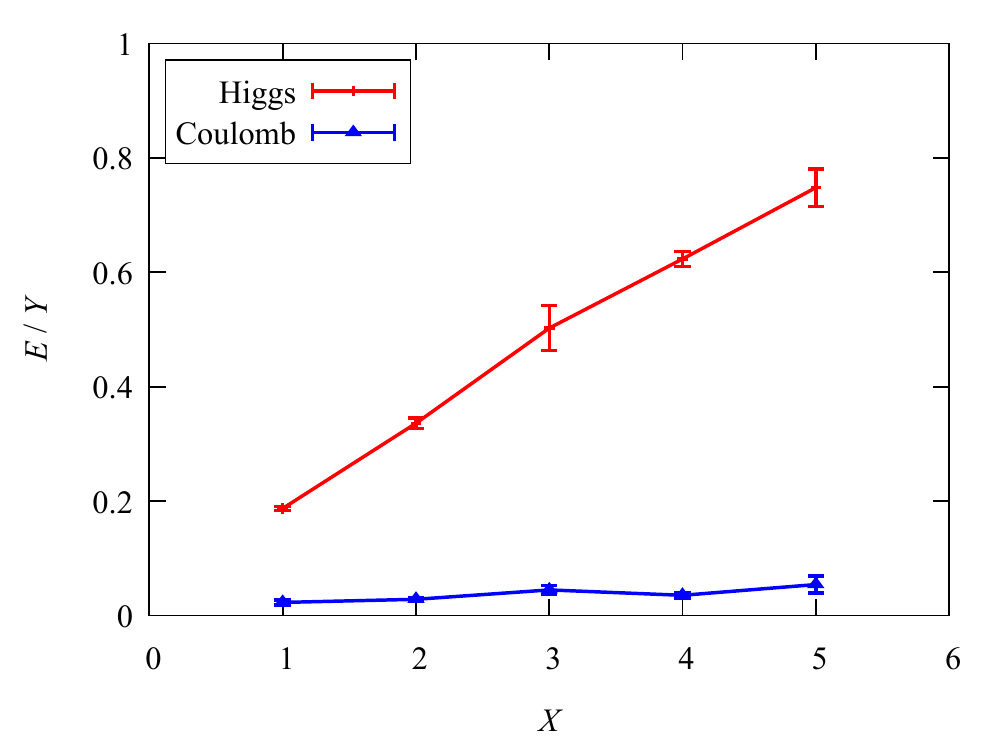}
\caption{
\label{figE}
Intervortex potential per length in the charge-2 lattice Abelian Higgs model.
The potentials in the Higgs phase ($\lambda=4$) and in the Coulomb phase ($\lambda=1$) are shown.
}
\end{center}
\end{figure}

The simulation results are shown in Fig.~\ref{figE}.
Changing the scalar self-coupling constant, we calculated the potential in the Higgs phase and the Coulomb phase.
In the Higgs phase, the potential is linear.
The vortices with fractional magnetic charge are confined.
Thus, only the states with zero magnetic charge will appear at low energy.
This will be the common property both for lattice and continuous superconductors.
In the Coulomb phase, the potential is almost flat.
As the U(1) symmetry is unbroken, the dual variables are massive, so they cannot propagate to long range.
The interaction between the vortices is diminished.

The linear potential is equivalent to the volume-law scaling of the 't Hooft surface, which is the criterion for the confinement of magnetic vortices.
This is dual to the volume-law scaling of the Wilson surface, which is the criterion for the confinement of electric strings.
In this model, however, the volume law is not exactly satisfied.
If the linear potential were exactly correct, the potential energy would be very large at long distance.
A dynamical vortex-antivortex pair will be created in-between the original vortex and antivortex to lower the total energy.
This is called the surface breaking as the analog of the string breaking of the Wilson loop \cite{Hayata:2019rem}.
The surface breaking will happen when the potential energy reaches the mass of the dynamical vortex-antivortex pair.
The potential will be flat above a critical distance.
This is not seen in Fig.~\ref{figE}.
We need the long-distance analysis in larger lattice volume.

\section{Summary}
\label{secsum}

In this paper, we obtained the formula for the t' Hooft surface in the lattice gauge Higgs models with the Z$_N$ topological order.
The formula is simple; the Z$_N$ element is multiplied to the hopping terms inside the 't Hooft surface.
We performed the lattice simulation in the charge-2 Abelian Higgs model.
We found that the the vortices with fractional magnetic charge are confined in the Higgs phase, and thus they do not emerge in the physical spectrum.
This is consistent with our understanding of superconducting vortices.

The advantage of our formulation is that the 't Hooft surface is calculable in the original path integral.
It is also attainable in the dual path integral, in which all the variables are dualized.
However, the dual approach works only when the duality between the two theories is ensured, e.g., in the London limit of type-II superconductors.
Our formulation is, therefore, more useful for the quantitative analysis in the original theory.

\begin{acknowledgments}
A.~Y.~was supported by JSPS KAKENHI Grant Number 19K03841.
The numerical calculations were carried out on SX-ACE in Osaka University.
\end{acknowledgments}

\appendix
\section{DERIVATION OF THE FORMULA}
\label{appder}

Let us derive the formula~\eqref{eqexW}.
The derivation is parallel to Ref.~\cite{Ukawa:1979yv}.
We introduce a special notation in this appendix.
When a bond $b$ connects $x$ and $x+\hmu$, we write as $\phi_{\partial b^{(1)}} = \phi(x)$, $\phi_{\partial b^{(2)}} = \phi(x+\hmu)$, and $U_b=U_\mu(x)$.
The hopping part \eqref{eqShop} is rewritten as
\begin{equation}
  \label{eqShop_bond}
  S_{\rm hop} = -\sum_b \bigg\{ \phi^*_{\partial b^{(1)}} U_b^N \phi_{\partial b^{(2)}} + \phi^*_{\partial b^{(2)}} U_b^{-N} \phi_{\partial b^{(1)}} \bigg\}\,.
\end{equation}
The summation is taken over all bonds $b$.

The scalar field $\phi$ is uniquely decomposed into a Z$_N$ part and a U(1)\,/\,Z$_N$ part.
When $\arg\phi(x) \in (-\pi,\pi]$, the decomposition is given by
\begin{equation}
  \phi(x) = e^{i2\pi \alpha(x)/N}\varphi(x)
\end{equation}
where $\alpha(x)\in\{0,1,\cdots,N-1\}$ and $\arg\varphi(x) \in (-\pi/N,\pi/N]$.
For a bond $b$ between $x$ and $x+\hmu$, we define
\begin{equation}
  \Delta_b\alpha = \alpha(x+\hmu) - \alpha(x)\,.
\end{equation}
The hopping part~\eqref{eqShop_bond} is rewritten by
\begin{equation}
  S_{\rm hop}[\alpha,A,\varphi] = \sum_b\left( e^{i2\pi\Delta_b\alpha/N}R_b[A,\varphi] + {\rm c.c.}\right)
\end{equation}
where $R_b[A,\varphi]$ is the remaining part dependent on $b$ and ``c.c.''~represents the complex conjugation.
The path integral becomes
\begin{equation}
\begin{split}
  &Z_0 = \int DAD\phi \ e^{-S} \\
  &=\int DAD\varphi\ e^{-S_{\rm gauge}[A]-S_{\rm local}[\varphi]}\prod_x\sum_{\alpha(x)=0}^{N-1}\ e^{-S_{\rm hop}[\alpha,A,\varphi]}\,.
\end{split}
\end{equation}
With the identity relation for Z$_N$ elements
\begin{equation}
 \frac1N \sum_{l=0}^{N-1} e^{-i2\pi nl/N} = \delta_{n0}  \quad ({\rm mod}\, N)
,
\end{equation}
we can expand as
\begin{equation}
  e^{-S_{\rm hop}} = \prod_b\sum_{l_b=0}^{N-1}I_{l_b}e^{i2\pi l_b\Delta_b\alpha/N}
  \label{eqhop_action}
\end{equation}
with the expansion coefficients
\begin{equation}
  I_{l_b} = \frac{1}{N}\sum^{N-1}_{n_b=0}e^{-i2\pi l_bn_b/N} \exp \left( -e^{i2\pi n_b/N}R_b-{\rm c.c.} \right)\,.
\end{equation}
It follows that
\begin{equation}
\begin{split}
\prod_x\sum_{\alpha(x)=0}^{N-1}\ e^{-S_{\rm hop}} &= \prod_x\sum_{\alpha(x)=0}^{N-1} \prod_b\sum_{l_b=0}^{N-1}I_{l_b}e^{i2\pi l_b\Delta_b\alpha/N}
\\
&=N\sum_{\{l_b\}'}\prod_b I_{l_b} \,.
\end{split}
  \label{eqhop_partition}
\end{equation}
$\sum_{\{l_b\}'}$ denotes the summation over the configurations which satisfy
\begin{equation}
  \sum_{b\ni x}l_b=0 \quad ({\rm mod}\, N)
  \label{eqconstraint}
\end{equation}
for all $x$.
In terms of the Z$_N$-valued link variable
\begin{equation}
  \xi_b \equiv e^{i2\pi l_b/N}\,,
\end{equation}
Eq.~\eqref{eqconstraint} is rewritten as
\begin{equation}
  \prod_{b\ni x} \xi_b = 1\,.
  \label{eqconstraint_xi}
\end{equation}
Since bonds are dual to cubes, it is convenient to use the dual cube variables $\xi_{c^*} \equiv \xi_b$.
The constraint on the original lattice, Eq.~\eqref{eqconstraint_xi}, is identical with
\begin{equation}
  \prod_{c^*\in h^*(x)}\xi_{c^*} = 1
  \label{eqdual_constraint}
\end{equation}
where $h^*(x)$ is the four-dimensional hypercube dual to the site $x$.
In order to satisfy Eq.~\eqref{eqdual_constraint} we define the dual plaquette variable $\zeta_{p^*}$ by
\begin{equation}
  \xi_{c^*} \equiv \prod_{p^* \in c^*} \zeta_{p^*}\,.
  \label{eqdual_var}
\end{equation}
The summation of $l_b$ with the constraint~\eqref{eqconstraint} is replaced by the summation of the unconstrained variables $\zeta_{p^*}$.
The path integral in the dual representation is given by
\begin{equation}
  Z_0 = N\int DAD\varphi \ e^{-S_{\rm gauge}-S_{\rm local}}\sum_{\{\zeta_{p^*}\}}\prod_b I_{l_b}.
\end{equation}

We define the 't Hooft surface on the dual lattice by
\begin{equation}
  \tilde{V}[\mathcal{S}^*] \equiv \prod_{p^*\in \mathcal{S}^*} \zeta_{p^*}^{q}
\end{equation}
where $q \in \{0,1,\dots,N-1\}$.
$\mathcal{S}^*$ represents a two-dimensional closed surface on the dual lattice.
With the three-dimensional volume $\mathcal{V}^*$ s.t.~$\partial \mathcal{V}^* = \mathcal{S}^*$, it also reads
\begin{equation}
  \tilde{V}[\mathcal{S}^*] =  \prod_{c^*\in \mathcal{V}^*} \xi_{c^*}^q .
\end{equation}
The expectation value of $\tilde{V}[\mathcal{S}^*]$ is given by
\begin{equation}
\begin{split}
  \langle \tilde{V}[\mathcal{S}^*] \rangle &\equiv \frac{N}{Z_0}\int DAD\varphi \ e^{-S_{\rm gauge}-S_{\rm local}} \sum_{\{\zeta_{p^*}\}} \tilde{V}[\mathcal{S}^*]\prod_b I_{l_b} .
\end{split}
\end{equation}

We go backward to arrive at the formula~\eqref{eqexW}.
Introducing $B(\mathcal{V}^*)$, which is the subset of the bonds penetrating $\mathcal{V}^*$, and
\begin{equation}
  k_b = \left\{
  \begin{array}{ll}
    1 & (b \in B(\mathcal{V}^*))
    \\
     &
    \\
    0 & (b \notin B(\mathcal{V}^*))
  \end{array}
  \right.
,
\end{equation}
we can rewrite as
\begin{equation}
  \tilde{V}[\mathcal{S}^*] = \prod_{b\in B(\mathcal{V}^*)} \xi_{b}^q = \prod_{b} e^{i2\pi l_b k_bq/N}
\end{equation}
It follows from Eq.~\eqref{eqdual_var} that
\begin{equation}
\label{eqsurface_dual}
  \begin{split}
    \langle \tilde{V}[\mathcal{S}^*] \rangle &\equiv \frac{N}{Z_0}\int DAD\varphi \ e^{-S_{\rm gauge}-S_{\rm local}}
    \\
    &\quad \times \sum_{\{l_b\}'}\prod_b I_{l_b} e^{i2\pi l_b k_bq/N}
  \\
  &= \frac{1}{Z_0}\int DAD\varphi \ e^{-S_{\rm gauge}-S_{\rm local}}
    \\
    &\quad \times\prod_b\sum_{l_b=0}^{N-1}\prod_x\sum_{\alpha(x)=0}^{N-1}I_{l_b}e^{i2\pi l_b\Delta_b\alpha/N}e^{i2\pi l_b k_bq/N}
  \\
  &\equiv \frac{1}{Z_0}\int DAD\varphi \ e^{-S_{\rm gauge}-S_{\rm local}} \prod_x\sum_{\alpha(x)=0}^{N-1}e^{-S'_{\rm hop}}
\end{split}
\end{equation}
where
\begin{equation}
  S'_{\rm hop} = \sum_b\left( e^{i2\pi\Delta_b\alpha/N}e^{i2\pi k_bq/N}R_b[A,\varphi] + {\rm c.c.}\right)\,.
  \label{eqShop_prime}
\end{equation}
Equation~\eqref{eqShop_prime} is nothing but Eq.~\eqref{eqSprime}.
Therefore,
\begin{equation}
  \langle \tilde{V}[\mathcal{S}^*] \rangle = \frac{Z_{\mathcal{S}^*}}{Z_0} \,.
\end{equation}
This is the end of the proof of Eq.~\eqref{eqexW}.

\section{SIMULATION DETAILS}\label{appsim}

We performed the lattice simulation with the hybrid Monte Carlo method.
We analyzed two cases: the Higgs phase and the Coulomb phase.
The scalar self-coupling constant was set at $\lambda=4$ for the Higgs phase and $\lambda=1$ for the Coulomb phase.
The other parameters were fixed at $1/e^2=2$ and $v^2 = 0.5$.
The lattice volume is $N_x N_y N_z \times N_\tau = 10^3 \times 20$ and all boundary conditions are periodic.
The size of the 't Hooft surface is $X=\{1,2,\cdots,5\}$, $Y = 10$, and $T=\{1,2,\cdots,10\}$.
We obtained the energy $E$ by fitting the data with Eq.~\eqref{eqET} in a finite range of $T$.
We checked that the results are insensitive to the fitting range of $T$.

From Eq.~\eqref{eqexW}, we see that the overlap between $Z_{{\mathcal S}^*}$ and $Z_0$ is $\exp(-\Delta S)$.
The overlap exponentially decreases as the surface size increases.
When the surface size is large, the relevant configurations hardly appear in the Monte Carlo sampling.
This is called the overlap problem.
To overcome the overlap problem, the expectation value \eqref{eqexW} was decomposed into the $XYT$ pieces,
\begin{equation}
 \frac{Z_{\mathcal{S}^*}}{Z_0}
= \frac{Z_{XYT}}{Z_{XYT-1}} \frac{Z_{XYT-1}}{Z_{XYT-2}} \cdots \frac{Z_{1}}{Z_{0}}
= \prod_k \frac{Z_k}{Z_{k-1}}
,
\end{equation}
and each piece $Z_k/Z_{k-1}$ was independently computed by the Monte Carlo simulation \cite{deForcrand:2000fi}.
The index $k$ means the number of the sites where $-1$ is multiplied to the hopping terms, so $Z_{\mathcal{S}^*} = Z_{XYT}$.

\bibliographystyle{apsrev4-1}
\bibliography{paper}

\end{document}